\documentclass[
article,
twocolumn,
superscriptaddress,
amsmath,amssymb,
prl,
]{revtex4-2}
\usepackage{graphicx}
\usepackage{hyperref}
\usepackage{dcolumn}
\usepackage{bm}
\usepackage{epstopdf}
\usepackage{romannum}
\usepackage{csquotes}
\usepackage[dvipsnames]{xcolor}
\usepackage{hyperref}
\usepackage{float}
\usepackage{graphicx}
\usepackage{fancyhdr}
\newcommand{\titlecase}[1]{%
	\MakeUppercase{#1}\MakeLowercase{\expandafter\lowercase\expandafter{\substring{#1}{2}{9999}}}}

\usepackage{lineno}

\begin{document}

\preprint{Boost of critical current density near quantum critical points in FeSe-Based superconductors with two superconducting domes}
\title{Boost of critical current density near quantum critical points in FeSe-Based superconductors with two superconducting domes}

\author{Wei Wei}
\affiliation{Key Laboratory of Quantum Materials and Devices of Ministry of Education, School of Physics, Southeast University, Nanjing 211189, China}

\author{Qiang Hou}
\affiliation{Key Laboratory of Quantum Materials and Devices of Ministry of Education, School of Physics, Southeast University, Nanjing 211189, China}

\author{Jiajia Feng}
\affiliation{Key Laboratory of Quantum Materials and Devices of Ministry of Education, School of Physics, Southeast University, Nanjing 211189, China}
\affiliation{Center for High Pressure Science and Technology Advanced Research (HPSTAR), Beijing 100094, P. R. China}

\author{Xinyue Wang}
\affiliation{Key Laboratory of Quantum Materials and Devices of Ministry of Education, School of Physics, Southeast University, Nanjing 211189, China}

\author{Xin Zhou}
\affiliation{Key Laboratory of Quantum Materials and Devices of Ministry of Education, School of Physics, Southeast University, Nanjing 211189, China}

\author{Nan Zhou}
\affiliation{Key Laboratory of Quantum Materials and Devices of Ministry of Education, School of Physics, Southeast University, Nanjing 211189, China}
\affiliation{Key Laboratory of Materials Physics, Institute of Solid State Physics,	HFIPS, Chinese Academy of Sciences, Hefei, 230031, China}
\author{Yan Meng}
\affiliation{Key Laboratory of Quantum Materials and Devices of Ministry of Education, School of Physics, Southeast University, Nanjing 211189, China}
\affiliation{School of Physical Science and Intelligent Engineering, Jining University, Qufu 273155, People’s Republic of China}

\author{Wei Zhou}
\affiliation{School of Electronic and Information Engineering, Changshu Institute of Technology, Changshu 215500, People’s Republic of China}

\author{Wenjie Li}
\affiliation{Key Laboratory of Quantum Materials and Devices of Ministry of Education, School of Physics, Southeast University, Nanjing 211189, China}
\affiliation{Department of Applied Physics, The University of Tokyo, Tokyo 113-8656, Japan}

\author{Xiangzhuo Xing}
\affiliation{School of Physics and Physical Engineering, Qufu Normal University, Qufu 273165, China}

\author{Tsuyoshi Tamegai}
\affiliation{Department of Applied Physics, The University of Tokyo, Tokyo 113-8656, Japan}

\author{Yue Sun}
\email{Corresponding author:sunyue@seu.edu.cn}
\affiliation{Key Laboratory of Quantum Materials and Devices of Ministry of Education, School of Physics, Southeast University, Nanjing 211189, China}

\author{Zhixiang Shi}
\email{Corresponding author:zxshi@seu.edu.cn}
\affiliation{Key Laboratory of Quantum Materials and Devices of Ministry of Education, School of Physics, Southeast University, Nanjing 211189, China}

\begin{abstract}
 \textbf{}
Recent studies have identified two superconducting domes in FeSe-based superconductors. It was discovered that each dome is accompanied by a distinct nematic quantum critical point (QCP): one associated with a pure nematic QCP, and the other with a nematic QCP entangled with antiferromagnetism (AFM). 
In this study, we delve into the evolution of the critical current density ($J_{\rm{c}}$) with doping in FeSe${_{1-x}}$(Te/S)${_{x}}$ single crystals, focusing on the behavior within the two superconducting domes. Surprisingly, three maxima of $J_{\rm{c}}$ were found in the two superconducting domes, with two sharp peaks in $J_{\rm{c}}$ observed precisely at the endpoints of the nematic phases, at $x$(Te) $\sim$ 0.5 for Te-doped and $x$(S) $\sim$ 0.17 for S-doped FeSe. The mechanisms of vortex pinning and the influence of quantum critical fluctuations have been extensively explored, emphasizing the contribution of quantum critical fluctuations in modulating $J_{\rm{c}}$. Additionally, an increase in $J_{\rm{c}}$ was also noted near FeSe$_{0.1}$Te$_{0.9}$, where its origin has been explored and discussed. This finding provides crucial clues about the existence of an ordered phase endpoint beneath the superconducting dome, offering an initial basis for further investigation into the potential presence of a QCP beneath it.
\\


\end{abstract}

\maketitle
\textbf{}
\pagenumbering{arabic}
\textbf{\begin{flushleft}
		1. Introduction
\end{flushleft}}
High-temperature superconductors (HTS) are quantum materials characterized by intriguing and complex phase diagrams, which include antiferromagnetic, ferromagnetic, nematic, superconducting, and strange metal phases emerging within different ranges of control parameters \( g \) (such as carrier concentration and pressure). The endpoint of ordered phase sometimes marks the formation of a quantum critical point (QCP) \cite{shibauchi2014quantum}. This is regarded as one of the most important breakthroughs in understanding the mechanisms of unconventional superconductivity. It also provides valuable insights into the origins of anomalous non-Fermi liquid behavior of and the coexistence of superconducting states with other exotic orders, such as magnetism, charge-density waves, and nematic states \cite{dai2012magnetism,pratt2009coexistence,christianson2009static,wang2011magnetic,iye2012gradual,iye2012microscopic,kasahara2010evolution,gruner2017charge}. A common paradigm in unconventional superconductors, including cuprates, iron-based compounds, and heavy fermion materials, is the presence of a QCP hidden beneath the superconducting dome. It is widely believed that quantum critical fluctuations strengthen the bosonic coupling strength, leading to a strong Cooper pairing in an usual way, thereby enhancing superconducting transition temperature ($ T_{\rm{c}} $) and forming a superconducting dome centered around the QCP \cite{shibauchi2014quantum}. However, in addition to the established relationship between QCP and $ T_{\rm{c}} $, recent investigations suggest that quantum critical fluctuations may also play a significant role in influencing the critical current density ($ J_{\rm{c}} $) \cite{jung2018peak,liu2022peak,hecher2018direct,wang2024quantum,tallon1999critical,naqib2019possible}. 
\begin{figure*}\center
	\includegraphics[width=1\linewidth]{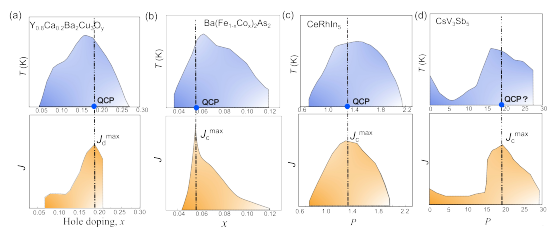}
	\caption{(a-d) $ T_{\rm{c}} $ (top) and $ J_{\rm{c}} $ (bottom) as functions of doping concentration ($ x $) or pressure ($ P $) in Y$_{0.8}$Ca$_{0.2}$Ba$_{2}$Cu$_{3}$O$_{y}$ \cite{talantsev2014hole,tallon1999critical,naqib2019possible}, Ba(Fe$_{1-x}$Co$_{x}$)$_{2}$As$_{2}$ \cite{hecher2018direct}, heavy-fermion superconductor CeRhIn$_{5}$ \cite{jung2018peak}, and the kagome superconductor CsV$ _{3} $Sb$ _{5} $ \cite{wang2024quantum}. The dot-dashed lines indicate the locations of the QCP (or the endpoint of the ordered phase).
	}\label{}
\end{figure*}

Given the established connection between the QCP and its role in enhancing high-$ T_{\rm{c}} $ superconductivity, it has been reported that the QCP may also affect $ J_{\rm{c}} $ by enhancing the depairing critical current density $ J_{\rm{d}} $. As observed in the hole-doped high-$ T_{\rm{c}} $ superconductor Y$ _{0.8} $Ca$ _{0.2} $Ba$ _{2} $Cu$ _{3} $O$ _{y} $ (YBCO) \cite{naqib2019possible}, the depairing critical current density $ J_{\rm{d}} $ and $ J_{\rm{c}} $ exhibits a significant peak at the critical doping (level $ p $ = 0.19), precisely where the pseudogap disappears \cite{ramshaw2015quasiparticle}. The phenomenon has also been observed in other cuprates \cite{naamneh2014doping}. Similarly, in iron-based, heavy-fermion, and Kagome superconductors, a correlation between $ J_{\rm{c}}^{\rm{max}} $ and the QCP (or the endpoint of the ordered phase) has also been observed, as shown in Figs. 1(b)-1(d) \cite{hecher2018direct,jung2018peak,wang2024quantum}. Putzke $ et$ $al $., through detailed measurements of the upper and lower critical fields in BaFe$ _{2} $(As$ _{1-x} $P$ _{x} $)$_{2} $, proposed that vortex core energy is enhanced near the QCP at $ x $ = 0.3 \cite{putzke2014anomalous}, and their findings align with the observed enhancement in $ J_{\rm{c}} $ \cite{ishida2018effects}. Therefore, investigating the potential impact of quantum critical fluctuations on $ J_{\rm{c}} $ could provide new insights into the role of the QCP in determining the performance of superconductors.

In this context, a system that exhibits the decoupling of \( T_{\rm{c}}^{\rm{max}} \) and QCP provides a good platform for investigating the relationship between $J_{\rm{c}}$ and QCP. FeSe-based superconductors undergo a structural transition from tetragonal to orthorhombic at $ T_{\rm{s}} \sim $ 90 K, accompanied by the nematic phase \cite{hsu2008superconductivity,shibauchi2020exotic}. The substitution of Se sites by isovalent sulfur \cite{yi2021hydrothermal,reiss2017suppression} or tellurium \cite{terao2019superconducting} is an effective method for adjusting the superconductivity and nematicity. Fig. 2(a) shows the complete superconducting phase diagram of FeSe\(_{1-x}\)(Te/S)\(_x\) single crystals, which has been previously established in our previous work \cite{QiangHou2024,sun2016electron}. Elastoresistivity measurements \cite{ishida2022pure} confirmed the presence of two nematic phase endpoints in FeSe-based superconductors. Subsequently, characteristics at the QCP, such as the divergence of the effective mass and the crossover from Fermi-liquid to non-Fermi-liquid behavior, have also been observed \cite{licciardello2019electrical}. Thus, the decoupled behavior of \( T_{c}^{\rm{max}} \) and QCP in FeSe-based superconductors offers a unique opportunity to explore the fundamental connection between $J_{\rm{c}}$ and QCP. Notably, as depicted in Fig. 2(b), we indeed observe that $J_{\rm{c}}$ in FeSe-based superconductors exhibits an enhancement at the putative nematic QCPs. The mechanisms of vortex pinning and the impact of quantum critical fluctuations
\begin{figure}[H]\center
	\includegraphics[width=1\linewidth]{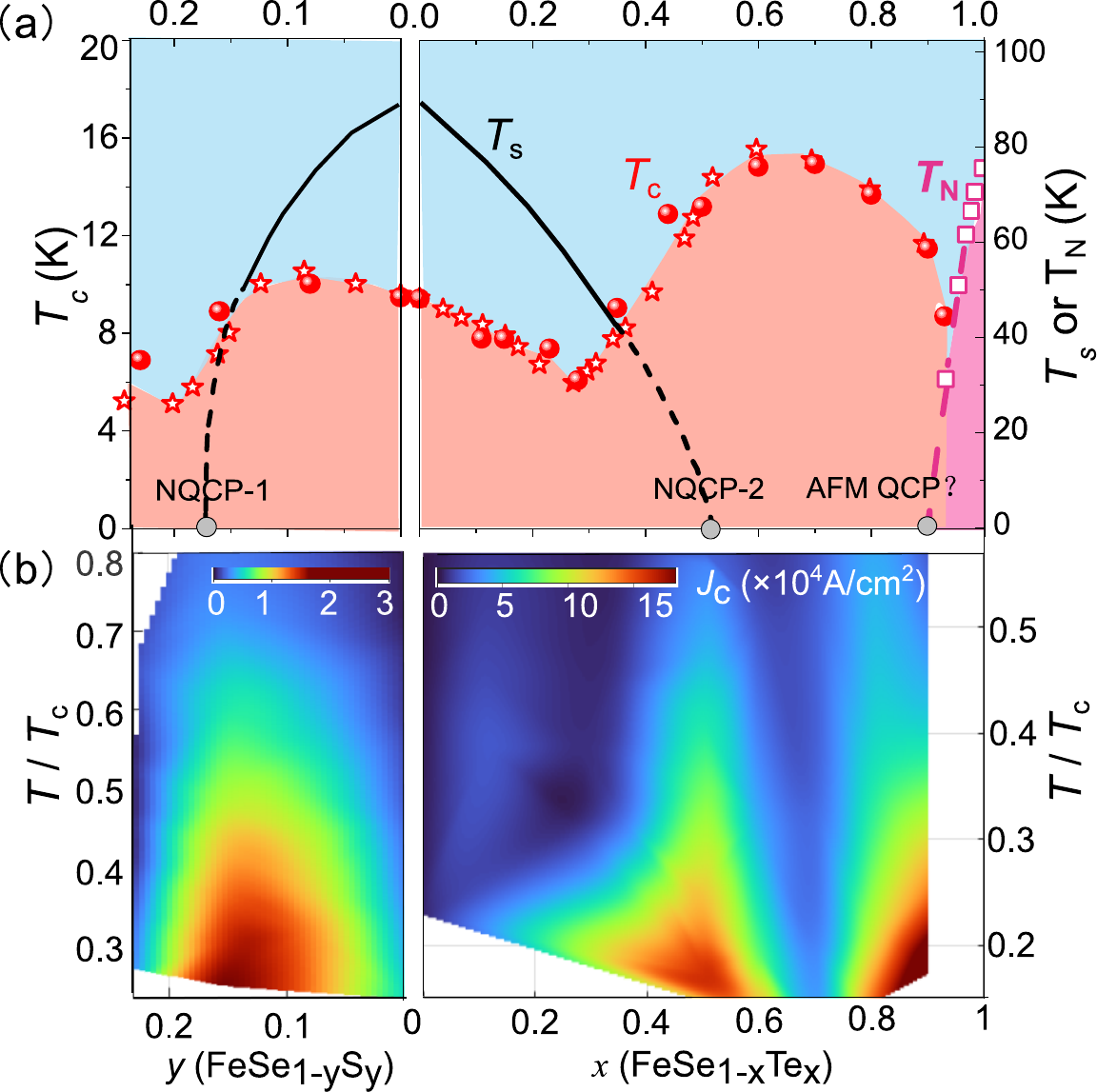}
	\caption{(a) The complete phase diagram of FeSe$ _{1-x} $(Te/S)$ _{x} $ single crystals. Black lines, solid circles, and squares correspond to $ T_{\rm{s}} $, $ T_{\rm{c}} $, and $ T_{\rm{N}} $, respectively. The open stars represent the $ T_{\rm{c}} $($ x $) values from Ref.\cite{mukasa2023enhanced}. (b) Contour plot of $ J_{\rm{c}} $ for FeSe$ _{1-x} $(Te/S)$ _{x} $ at a normalized temperature under self-field ($ H $ = 0 T). The color represents the scale of the critical current density. The $ J_{\rm{c}}$ at zero magnetic field for FeSe$ _{0.96} $S$ _{0.04} $ is derived from the study of Mahmoud Abdel-Hafiez $ et$ $al$. \cite{abdel2015superconducting}.
	}\label{}
\end{figure}

\begin{figure*}\center
	\includegraphics[width=1\linewidth]{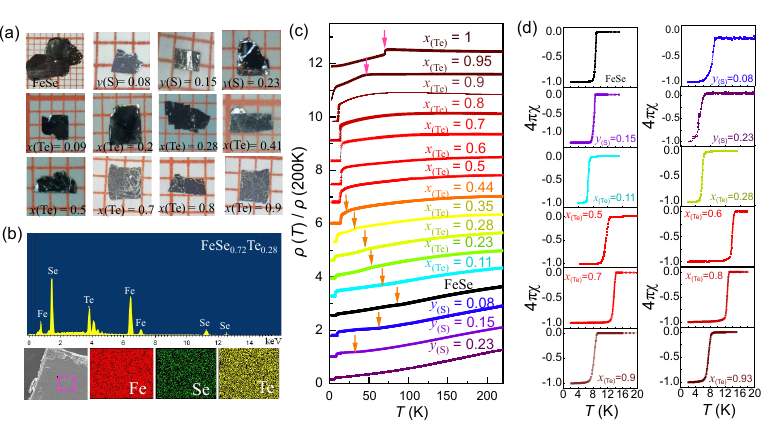}
	\caption{(a) Images of FeSe$_{1-x}$(Te/S)$_{x}$ single crystals; (b) The EDX spectrum (top panel) and compositional mappings (bottom panels) of Fe, Se, and Te within the selected rectangular region are shown for a representative FeSe$ _{0.72} $Te$ _{0.28} $ single crystal; 
		(c) Temperature dependence of the normalized resistivity, $ \rho $($ T $)/$ \rho $(200 K), for different FeSe$_{1-x}$(Te/S)$_{x}$ single crystals. The orange and pink arrows indicate the nematic ($ T_{\rm{s}} $) and antiferromagnetic transitions ($ T_{\rm{N} }$), respectively;
		(d) Temperature dependences of ZFC magnetizations at 5 Oe parallel to $ c $-axis for FeSe$_{1-x}$(Te/S)$_{x}$ single crystals.
	}\label{}
\end{figure*}
\noindent
have been thoroughly discussed, highlighting the previously neglected role of quantum critical fluctuations in modulating $J_{\rm{c}}$. The relationship between the nematic QCPs and $ J_{\rm{c}} $ may allow $ J_{\rm{c}} $ to serve as a probe for determining the precise location of the QCP, or at least as a potential tool for identifying the vanishing point of the ordered phase beneath the superconducting dome.

\textbf{\begin{flushleft}
	2. Materials and Methods
\end{flushleft}}
High-quality single crystals of FeSe$ _{1-y} $S$ _{y} $ (0 $ \leq $ $ y $ $ \leq $ 0.23) and FeSe$ _{1-x} $Te$ _{x} $ (0 $ \leq $ $ x $ $ \leq $ 0.5) were grown using the chemical vapor transport (CVT) method \cite{sun2016electron,QiangHou2024}. Stoichiometric mixtures of Fe, Se, and Te/S were sealed in an evacuated quartz tube, with AlCl$ _{3} $ and KCl serving as transport agents. The temperatures of the source and sink sides were controlled at 500 and 300 $^\circ$C for 0 $ \leq $ $ x $ $ \leq $ 0.1 and 0 $ \leq $ $ y $ $ \leq $ 0.23 (650 and 250 $^\circ$C for 0.1 $ < $ $ x $ $ \leq $ 0.5), respectively. After more than 35 days, single crystals with dimensions in millimeters can be obtained in the cold end. Utilizing a combination of the self-flux method and Te-vapor annealing techniques, high-quality single crystals of FeSe$ _{1-x} $Te$ _{x} $  (0.5 $ < $ $ x $ $ \leq $ 1) were obtained, as previously described in our previous review paper \cite{sun2019review}. 

Elemental analysis was carried out using a scanning electron microscope equipped with an energy-dispersive x-ray (EDX) spectroscopy probe. Resistivities were measured using a standard four-probe technique with a Physical Property Measurement System (PPMS, Quantum Design) at temperatures as low as 2 K and magnetic fields up to 9 T. Magnetization measurements were performed using a commercial superconducting quantum interference device (SQUID) magnetometer (MPMS-XL5, Quantum Design).

\begin{figure*}\center
	\includegraphics[width=1\linewidth]{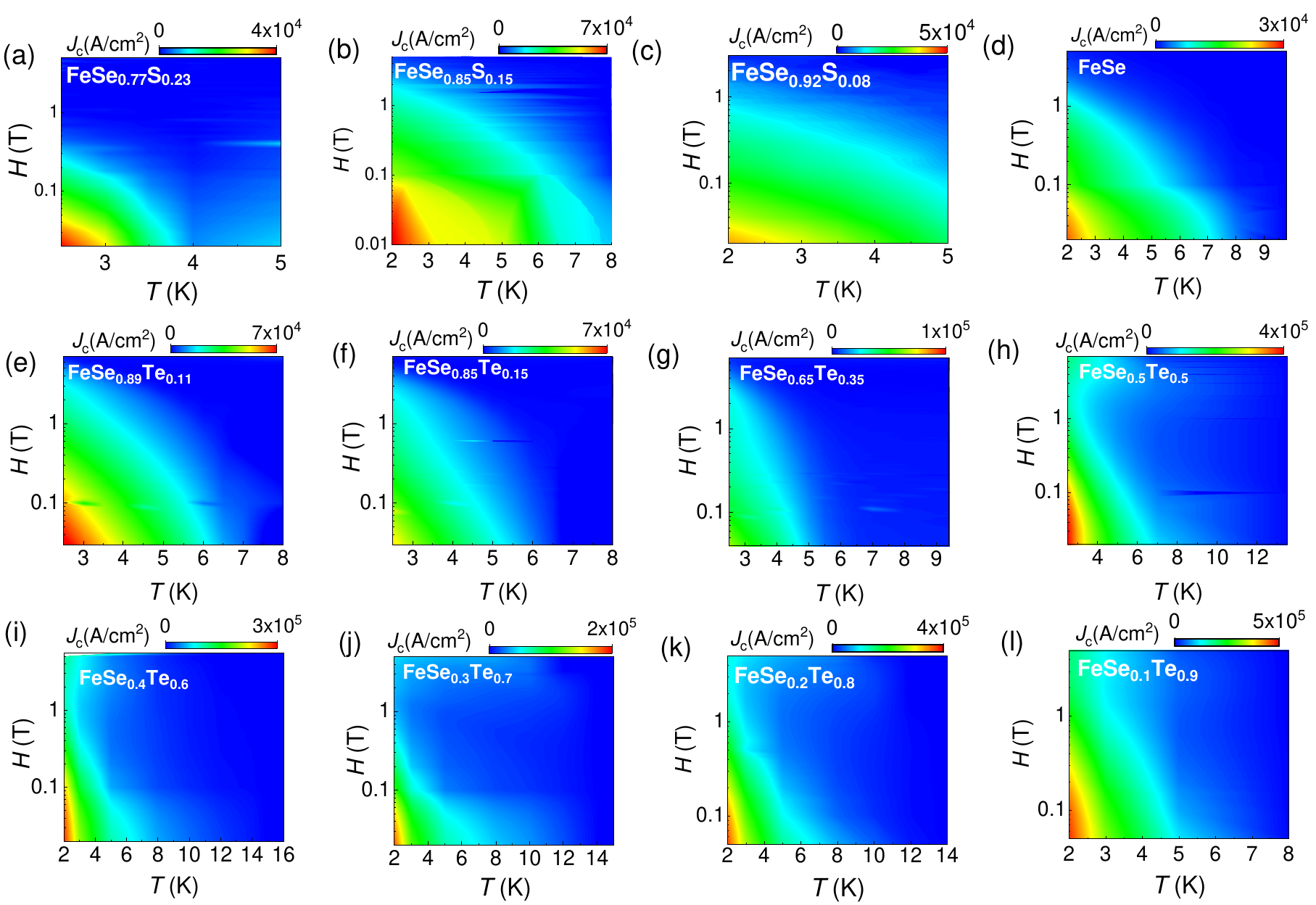}
	\caption{(a-i) The dependencies of $ J_{\rm{c}} $ on temperature ($T$) and magnetic field ($H$) for (a) FeSe$ _{0.77} $S$ _{0.23} $, (b) FeSe$ _{0.85} $S$ _{0.15} $, (c) FeSe$ _{0.92} $S$ _{0.08} $, (d) FeSe, (e) FeSe$ _{0.89} $Te$ _{0.11} $, (f) FeSe$ _{0.85} $Te$ _{0.15} $, (g) FeSe$ _{0.65} $Te$ _{0.35} $, (h) FeSe$ _{0.5} $Te$ _{0.5} $, (i) FeSe$ _{0.4} $Te$ _{0.6} $, (j) FeSe$ _{0.3} $Te$ _{0.7} $, (k) FeSe$ _{0.2} $Te$ _{0.8} $, and (l) FeSe$ _{0.1} $Te$ _{0.9} $ single crystals, respectively.}\label{}
\end{figure*}

\textbf{\begin{flushleft}
	3 Results and Discussion
\end{flushleft}}

The FeSe$ _{1-x} $Te$ _{x} $ (0 $ \leqslant x  \leqslant $ 1) and FeSe$ _{1-y} $S$ _{y} $(0 $ < $ $ y $ $ \leqslant $ 0.23) single crystals exhibit a shiny appearance, as shown in Figure 3(a) through optical images. The actual composition was determined using EDX. Fig. 3(b) presents the EDX spectrum (top panel) and compositional mappings (bottom panels) of Fe, Se and Te within the selected rectangular region for a representative FeSe$ _{0.72} $Te$ _{0.28} $ single crystal, indicating the homogeneous distribution of Fe, Se and Te. The evolution of the temperature-dependent normalized resistivity, $\rho(T)/\rho(200$ K), with different S/Te contents, is shown in Fig. 3(c). The curves have been vertically shifted for better visibility and clarity. The orange arrows indicate the nematic transition $ T_{\rm{s}} $, defined as the peak position in the temperature derivative of $ \rho (T)$/$ \rho $(200 K) (shown in Supplementary Figure S1 \cite{SM}). As the Te/S content increases, the $ T_{\rm{s}} $ gradually decreases and eventually becomes unobservable when the system enters the superconducting state. Previous research observed the nematic QCPs at the Te $ \sim $ 0.5 and S $ \sim $ 0.17 through the measurement of elastoresistivity \cite{ishida2022pure}. For FeSe$ _{1-x} $Te$ _{x} $ (0.95 $ \leqslant $ $ x $ $ < $ 1), an obvious AFM transition can be observed, as marked by the solid magenta arrows, consistent with previous reports \cite{sun2016influence}. Fig. 3(d) shows the temperature dependence of the zero-field-cooled (ZFC) magnetization under an applied magnetic field of 5 Oe with $ H$ $\Arrowvert$ $c $. Clearly, the magnetization measurement exhibits a sharp superconducting transition, and in alignment with the resistivity measurement. After considering the demagnetization effect, the shielding volume fraction approaches $ \sim $ 100 $ \% $ at 2 K, confirming the bulk nature of the superconductivity.  

$ J_{\rm{c}} $ of FeSe$_{1-x}$(Te/S)$_{x}$ single crystals was measured using magnetic hysteresis loops (MHLs) at various temperatures for $H$ $\Arrowvert$ $c$. Typical results for FeSe$ _{0.77} $S$ _{0.23} $, FeSe$ _{0.85} $S$ _{0.15} $, FeSe$ _{0.92} $S$ _{0.08} $, FeSe, FeSe$ _{0.89} $Te$ _{0.11} $, FeSe$ _{0.85} $Te$ _{0.15} $, FeSe$ _{0.65} $Te$ _{0.35} $, FeSe$ _{0.5} $Te$ _{0.5} $, FeSe$ _{0.4} $Te$ _{0.6} $, FeSe$ _{0.3} $Te$ _{0.7} $, FeSe$ _{0.2} $Te$ _{0.8} $, and FeSe$ _{0.1} $Te$ _{0.9} $ are shown in Supplementary Figures S2(a)–(l) \cite{SM}, respectively. The magnetization decreases as the magnetic field increases, and symmetric hysteresis loops are observed, indicating  the dominance of the bulk pinning. The field dependence of $ J_{\rm{c}} $ was calculated from the MHLs using the Bean model \cite{bean1964magnetization}:
\begin{eqnarray*}
	\\J_{c} =20\dfrac{\Delta M}{a(1-a/3b)},
\end{eqnarray*}
where $\Delta M $ represents the difference between the magnetization values ($ M_{\rm {down}} $ - $ M_{\rm{up}} $) measured when sweeping the fields up and down. $ M_{\rm{up}} $ [emu cm$ ^{-3} $] and $ M_{\rm{down}} $ [emu cm$ ^{-3} $] are the magnetization values during upward and downward field sweeps, respectively. Additionally, $ a $ [cm] and $ b $ [cm] denote the sample widths, with $ a $ being smaller than $ b $. The obtained $ J_{\rm{c}} $ at different temperatures are depicted in Supplementary Figure S3 \cite{SM}. Contour plots in Figure 4 illustrate the dependencies of $ J_{\rm{c} }$ on temperature and magnetic field. In this plot, red and blue colors correspond to high and low $ J_{\rm{c} }$, respectively. As the temperature and magnetic field increases, $ J_{\rm{c} }$ exhibit a generally decreasing trend. 
\begin{figure}\center
	\includegraphics[width=1\linewidth]{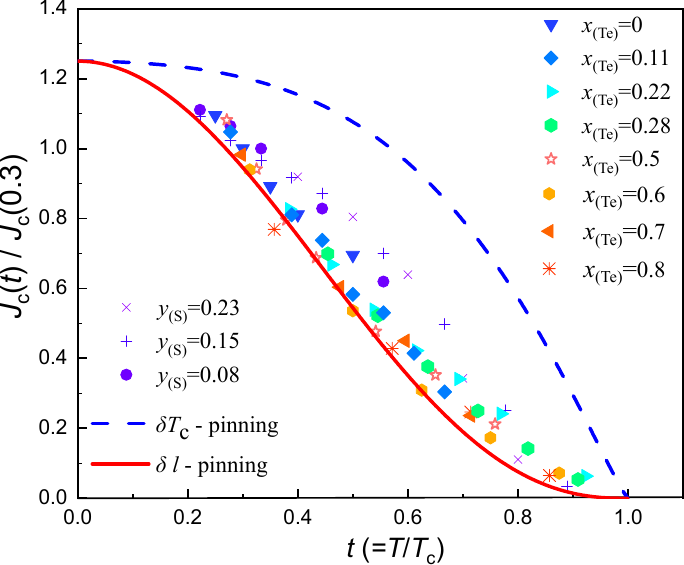}
	\caption{$ J_{\rm{c}}(t)$ normalized by $ J_{\rm{c}} $($ t $=0.3) as a function of the reduced temperature $ t $ = $ T $/$ T_{\rm{c}} $ for all the crystals under zero magnetic field. The normalized $ J_{\rm{c}} $($ t $) collapse onto a universal curve. Blue dashed line and red solid curve are for $ \delta T_{\rm{c}} $-pinning and $ \delta l $-pinning, respectively.
	}\label{}
\end{figure}
To gain deeper insights into the behavior of $ J_{\rm{c} }$, we first performed a pinning analysis. In Figure 5, we present $J_\mathrm{c} (t)$/$J_\mathrm{c} (t=0.3)$ as a function of the normalized temperature $ t $ = $ T $/$ T_{\rm{c} } $ for all the crystals under zero magnetic field. 
It is evident that the normalized $J_\mathrm{c}(t)$ roughly collapses onto an universal curve. This results proved that the pinning mechanism is not changed between crystals. 

Griessen $ et $ $ al $. proposed a model suggesting that two primary pinning mechanisms can describe the disorder in the order parameter introduced by local variations in either the superconducting transition temperature ($ \delta T_{\rm{c}}$) or the mean free path ($ \delta l$) of a material \cite{griessen1994evidence}.  
For $ \delta l$ pinning and $ \delta T_{\rm{c} } $ pinning, $ J_{\rm{c} }$ follows the following laws, respectively \cite{xiang2013evidence,griessen1994evidence}:
\begin{eqnarray*}
	\\J_{\rm{c}} (t)=J_{\rm{c}}(0)(1-t^{2})^{5/2}(1+t^{2})^{-1/2},
\end{eqnarray*}
and
\begin{eqnarray*}
	\\J_{\rm{c}} (t)=J_{\rm{c}}(0)(1-t^{2})^{7/6}(1+t^{2})^{5/6}.
\end{eqnarray*}
Fig. 5 shows the fitting results by both $\delta T_\mathrm{c}$ and  $\delta l$ pinning. Clearly, the pinning mechanism is different from the conventional model for $\delta T_\mathrm{c}$-pinning (blue dashed line) in the whole temperature region. The normalized data follows close to the $\delta l$ pinning curve (red solide line), yet it cannot be precisely matched by the $\delta l$ pinning mechanism. Similar results have also been observed in IBSs ``122" system \cite{vlasenko2015unconventional}. This is due to the derivation of this model is restricted to the weak collective pinning theory, making it unable to capture the behavior of most IBSs containing both strong and weak pinnings \cite{blatter1994vortices,griessen1994evidence}. A more comprehensive study of the vortex pinning behavior will be presented in our next work. 

We present $ J_{\rm{c}} $ for FeSe$ _{1-x} $(Te/S)$ _{x} $ under normalized magnetic fields ($ H /H_{\rm{c2}}$) at normalized temperatures ($ T / T_{\rm{c} }$) in Fig. S5. In Figs. S5(a) and S5(b), the magnitude of $ J_{\rm{c}} $ is indicated by the color scale. The upper critical field $H_{\rm{c2}}$ was determined by using PPMS, with the temperature-dependent resistivity curves under various magnetic fields shown in Fig. S4. The higher $H_{\rm{c2}}$ data were derived from the report by K. Mukasa $ et$ $al. $ \cite{mukasa2023enhanced}. The variation of $ J_{\rm{c}} $ observed in the normalized magnetic field and temperature is surprising. To show the variation of $ J_{\rm{c}} $ more clearly and illustrate the relationship between the $ J_{\rm{c}} $ and QCP, two phase diagrams were constructed, incorporating $ T_{\rm{c} }$, the structural (nematic) transition temperature $ T_{\rm{s} }$, the Néel temperature $ T_{\rm{N} }$, and a contour plot of the $ J_{\rm{c} }$ at zero magnetic field and normalized temperature ($ T $/$ T_{\rm{c} }$). The peak of $ J_{\rm{c}} $ occurs precisely at the point of the disappearance of the ordered phase, as determined by elastoresistivity measurements. Although the endpoint of the AFM phase has not been reported, it is surprisingly observed that the enhancement of $ J_{\rm{c}} $ occurs at the endpoint of the extension of the AFM ordered phase. 
Quantum fluctuations have long been proposed as a potential pairing glue mediating the formation of superconducting Cooper pairs in strongly correlated electron systems \cite{shibauchi2014quantum}. Therefore, it may play a significant role in modulating $ J_{\rm{c}} $, as has been observed in many unconventional superconductors. In cuprates, the divergence of effective mass $ m^{\ast} $ and the highest $ H_{\rm{c2}} $ at the point where the pseudogap disappears suggest the existence of a QCP in YBCO \cite{ramshaw2015quasiparticle}. The peak in $ J_{\rm{c}} $ does not align with the maximum $ T_{\rm{c}} $ but instead occurs at the QCP \cite{talantsev2014hole,tallon1999critical,naqib2019possible}, implying a contribution from the quantum critical fluctuation. 
And the magnitude of the peak in zero-field $ J_{\rm{c} }$ is believed to be primarily influenced by the quantum critical fluctuations associated with the hidden AFM QCP and non-magnetic QCP in CeRhIn$ _{5} $ and (Ca$ _{x} $Sr$ _{1-x} $)$ _{3} $Rh$ _{4} $Sn$ _{13} $ \cite{jung2018peak,liu2022peak}. 
The observed correlation between QCPs and $ J_{\rm{c}} $ also appears in iron-chalcogenide superconductors. Systematic elastoresistivity measurements reveal signatures of nematic QCPs in FeSe$ _{1-x} $(Te/S)$ _{x} $. Numerous studies 
further support the existence of a nematic QCP through characteristic signatures including diverging effective mass and a crossover from Fermi-liquid to non-Fermi-liquid behavior \cite{licciardello2019electrical}. The parameter $\mu_{0}H_{\rm{Pauli}}/T_{\rm{c}}$ reaches its maximum near FeSe$ _{0.52} $Te$ _{0.48} $, providing evidence for enhanced electron pairing strength \cite{mukasa2023enhanced}. This enhancement aligns with theoretical frameworks suggesting nematic quantum fluctuations as effective mediators of Cooper pairing \cite{kontani2010orbital,lederer2015enhancement,lederer2017superconductivity}. Therefore, the universal enhancement of $J_{\rm{c}}$ across both pure nematic QCP and antiferromagnetic intertwined nematic QCP in FeSe$ _{1-x} $(Te/S)$ _{x} $ systems provides evidence for the correlation between quantum critical fluctuations and $ J_{\rm{c}} $. The confirmation of this relationship may enable $J_{\rm{c}}$ to serve as a probe for identifying QCPs beneath the superconducting dome, yet more  research is needed to confirm this relationship. 
Additionally, apart from the peaks of $ J_{\rm{c}} $ observed at nematic QCPs, we also noticed a significant enhancement of $ J_{\rm{c}} $ near FeSe$ _{0.1} $Te$ _{0.9} $. Interetingly, the location of this maximum $ J_{\rm{c}} $ happens to be near the Te concentration where $ T_{\rm{N}} $ is extrapolated to 0 K. According to the results in heavy fermion system, it indicates that there may be an AFM QCP leading to the enhancement of $ J_{\rm{c}} $ \cite{jung2018peak}. In fact, near $ x \sim$  0.9, we also observed a striking deviation from conventional Fermi liquid behavior in the temperature dependence of normal-state resistivity \cite{QiangHou2024}. This observation provides a foundation for further investigation into whether an AFM QCP or an AFM endpoint exists beneath the superconducting dome. 

Of course, we cannot totally exclude the effect of pinning, mainly from the structure domains, on the peak of $ J_{\rm{c}} $. However, until now, there is still no evidence that the amount of structure domains reaches a maximum at nematic QCPs in FeSe$ _{1-x} $(Te/S)$ _{x} $. Furthermore, systematic investigations on both twinned and detwinned Ba(Fe,Co)$_{2}$As$_{2}$ single crystals have clearly demonstrated that twin boundaries are not responsible for the maximum $ J_{\rm{c}} $ \cite{hecher2018direct}. Therefore, we speculate that the effect of pinning is not the main cause of the $ J_{\rm{c}} $ peak in FeSe$ _{1-x} $(Te/S)$ _{x} $.

\textbf{\begin{flushleft}
4 Conclusions
\end{flushleft}}
In summary, we synthesized FeSe$ _{1-x} $(Te/S)$ _{x} $ single crystals, establishing comprehensive $T_{\rm{c}}$ and $J_{\rm{c}}$ phase diagrams. Through the simple method distinct from elastoresistivity measurements, we reveal a significant enhancement of $J_{\rm{c}} $ as the system is tuned towards the two different nematic QCPs. The mechanisms of vortex pinning and the influence of quantum critical fluctuations have been extensively explored.
Universal enhancement of $J_{\rm{c}} $ observed at different QCPs, one associated with a pure nematic QCP and the other with a nematic QCP entangled with AFM, implies the crucial role of quantum critical fluctuations in modulating $J_{\rm{c}} $.
The establishment of this relationship provides clues regarding the existence of the QCP beneath the superconducting dome.
\\

\textbf{\begin{flushleft}
		CRediT authorship contribution statement 
\end{flushleft}}
\textbf{Wei Wei} and \textbf{Qiang Hou} contributed equally to this work. \textbf{Wei Wei} and \textbf{Qiang Hou}: Conceptualization, Data Curation, Formal analysis, and Writing-Original draft preparation. \textbf{Jiajia Feng}, \textbf{Xinyue Wang, Xin Zhou, Nan Zhou, Yan Meng, Wei Zhou, Wenjie Li}, and \textbf{Xiangzhuo Xing}: Methodology. \textbf{Tsuyoshi Tamegai}: Supervision, Writing-Review $\&$ Editing. \textbf{Yue Sun} and \textbf{Zhixiang Shi}: Formal analysis, Conceptualization, Funding
acquisition, Supervision, Project administration.

\textbf{\begin{flushleft}
		Declaration of competing interest 
\end{flushleft}}
The authors declare that they have no known competing financial interests or personal relationships that could have appeared to influence the work reported in this paper.

\textbf{\begin{flushleft}
		Data availability 
\end{flushleft}}
Data will be made available on request.

\textbf{\begin{flushleft}
		Acknowledgements
\end{flushleft}}

This work was partly supported by the National Key R$\& $D Program of China (Grant No. 2018YFA0704300, No. 2024YFA1408400), the National Natural Science Foundation of China (Grants No. 12374136, No. 12374135, No. 12204487), the open research fund of Key Laboratory of Quantum Materials and Devices (Southeast University), and the Natural Science Foundation of Shandong Province (No. ZR2022QA040).


\bibliographystyle{unsrt}
\bibliography{ref}

\pagebreak

\onecolumngrid
\newpage
\setcounter{equation}{0}
\setcounter{figure}{0}
\setcounter{table}{0}

\makeatletter
\renewcommand{\theequation}{S\arabic{equation}}
\renewcommand{\thefigure}{S\arabic{figure}}
\begin{figure*}\center
	\includegraphics[width=0.7\linewidth]{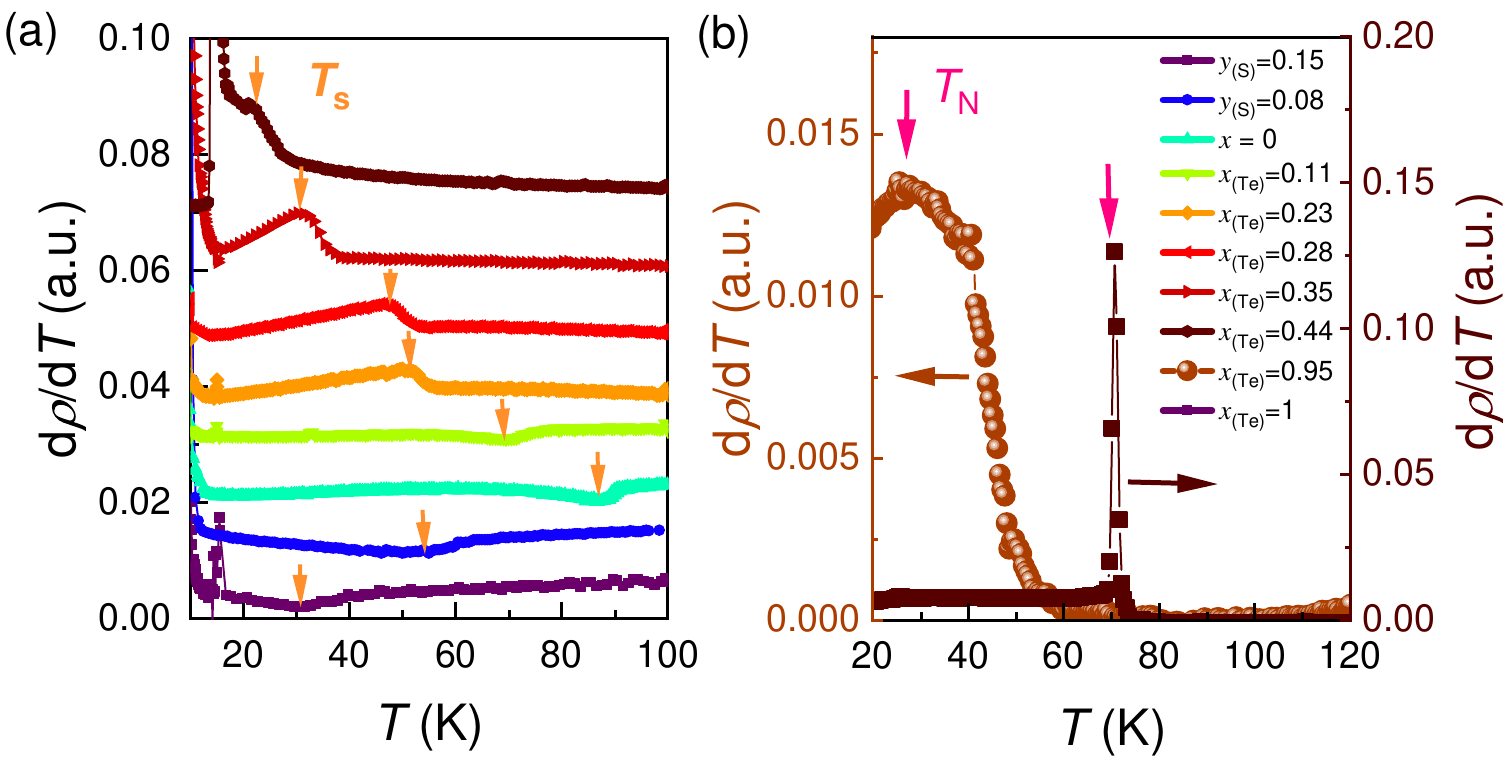}
	\caption{The first derivative of $ \rho – T $ for FeSe$ _{1-x} $(Te/S)$ _{x} $. The orange and pink arrows indicate the nematic ($ T_{\rm{s}} $) and antiferromagnetic transitions ($ T_{\rm{N} }$), respectively.}\label{}
\end{figure*}
\begin{figure*}\center
	\includegraphics[width=1\linewidth]{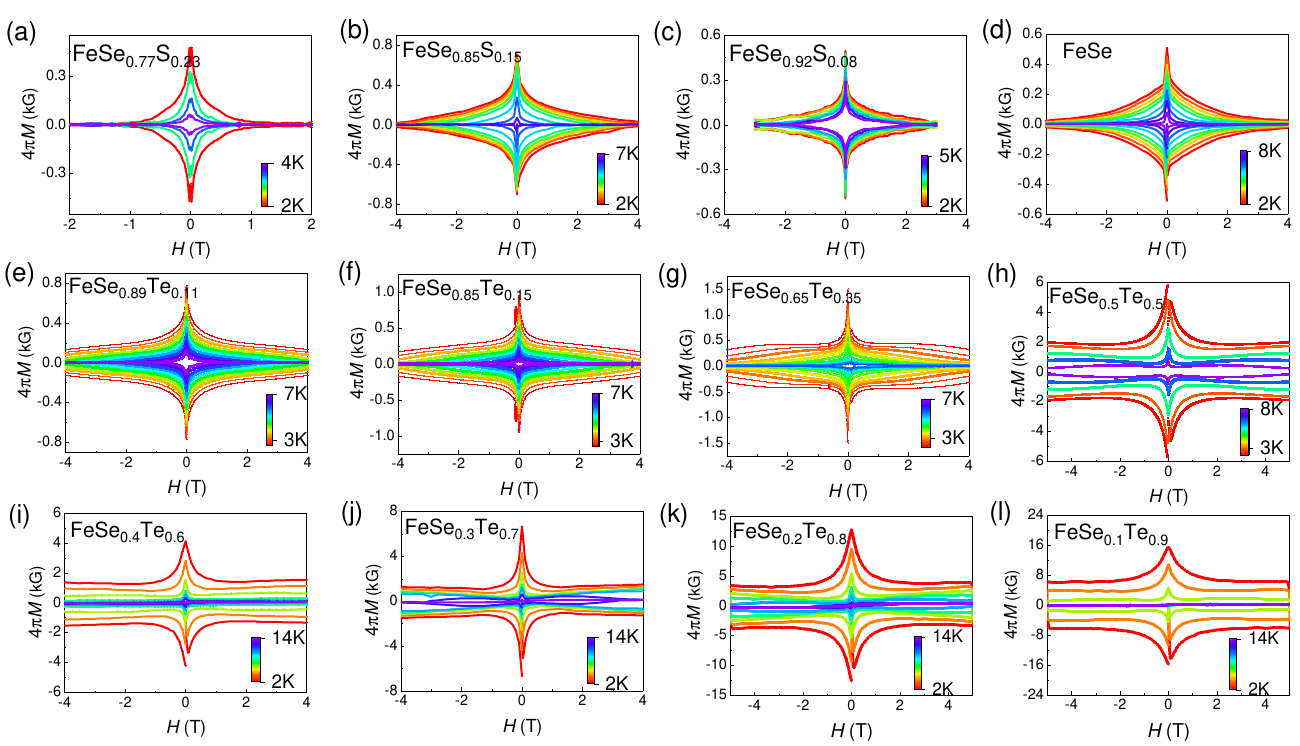}
	\caption{Magnetic hysteresis loops of FeSe$ _{1-x} $(Te/S)$ _{x} $ single crystals at different temperatures for (a) FeSe$ _{0.77} $S$ _{0.23} $, (b) FeSe$ _{0.85} $S$ _{0.15} $
		(c) FeSe$ _{0.92} $S$ _{0.08} $, (d) FeSe, (e) FeSe$ _{0.89} $Te$ _{0.11} $, (f) FeSe$ _{0.85} $Te$ _{0.15} $, (g) FeSe$ _{0.65} $Te$ _{0.35} $, (h) FeSe$ _{0.5} $Te$ _{0.5} $, (i) FeSe$ _{0.4} $Te$ _{0.6} $, (j) FeSe$ _{0.3} $Te$ _{0.7} $, (k)FeSe$ _{0.2} $Te$ _{0.8} $, and (l) FeSe$ _{0.1} $Te$ _{0.9} $.}\label{}
\end{figure*}
\begin{figure*}\center
	\includegraphics[width=1\linewidth]{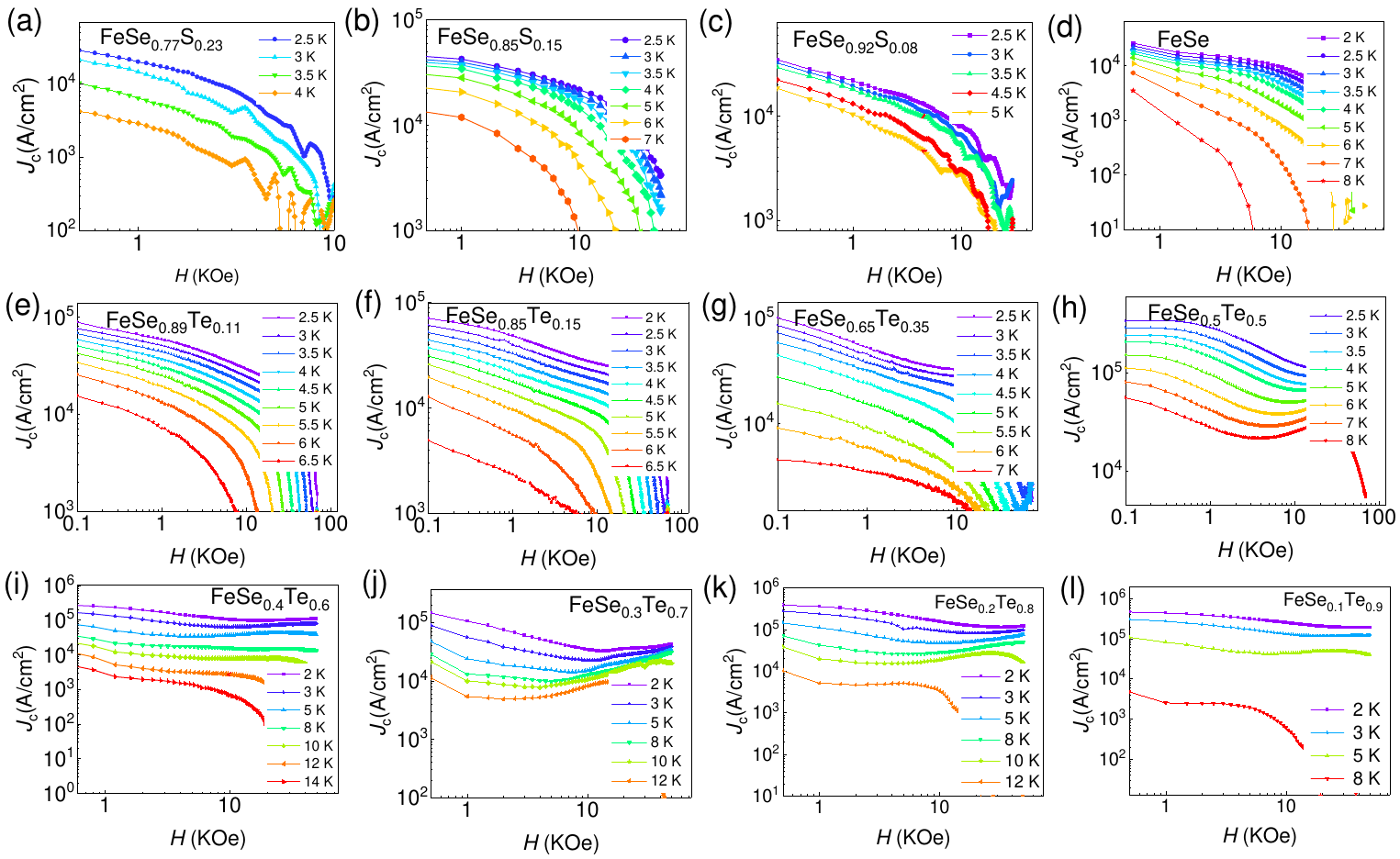}
	\caption{Magnetic field dependence of critical current densities with $ H$ $\Arrowvert$ $ c $ for (a) FeSe$ _{0.77} $S$ _{0.23} $, (b) FeSe$ _{0.85} $S$ _{0.15} $
		(c) FeSe$ _{0.92} $S$ _{0.08} $, (d) FeSe, (e) FeSe$ _{0.89} $Te$ _{0.11} $, (f) FeSe$ _{0.85} $Te$ _{0.15} $, (g) FeSe$ _{0.65} $Te$ _{0.35} $, (h) FeSe$ _{0.5} $Te$ _{0.5} $, (i) FeSe$ _{0.4} $Te$ _{0.6} $, (j) FeSe$ _{0.3} $Te$ _{0.7} $, (k)FeSe$ _{0.2} $Te$ _{0.8} $, and (l) FeSe$ _{0.1} $Te$ _{0.9} $.}\label{}
\end{figure*}

\begin{figure*}\center
	\includegraphics[width=1\linewidth]{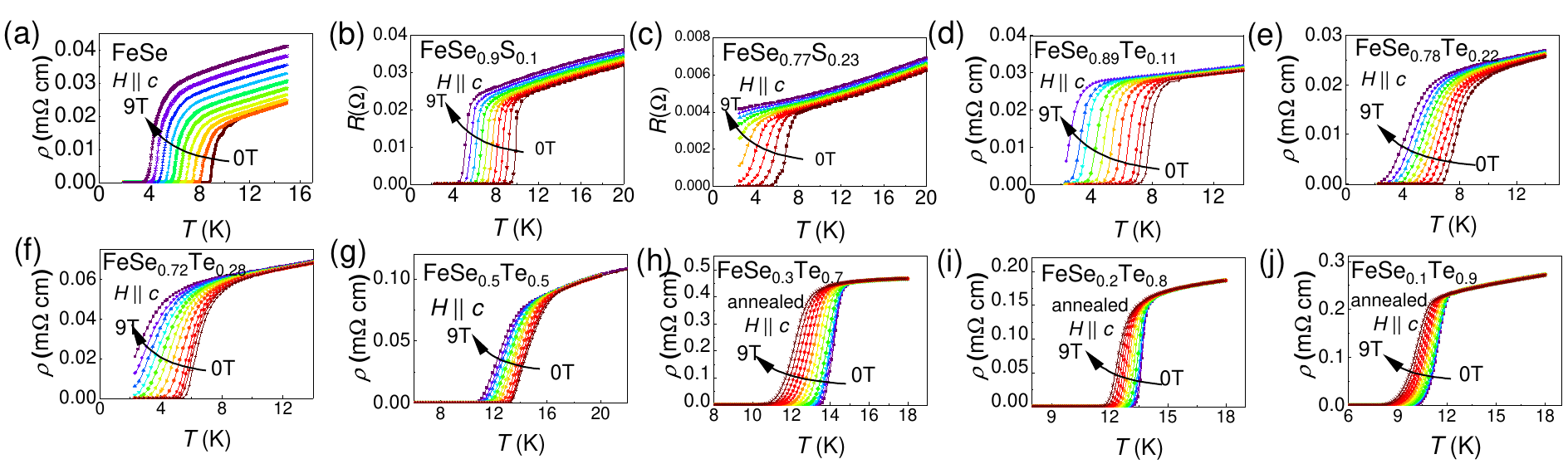}
	\caption{Temperature dependence of resistance for $ H$ $\Arrowvert$ $ c $ under magnetic fields from 0 to 9 T for FeSe$ _{1-x} $(Te/S)$ _{x} $. }\label{}
\end{figure*}

\begin{figure*}\center
	\includegraphics[width=0.8\linewidth]{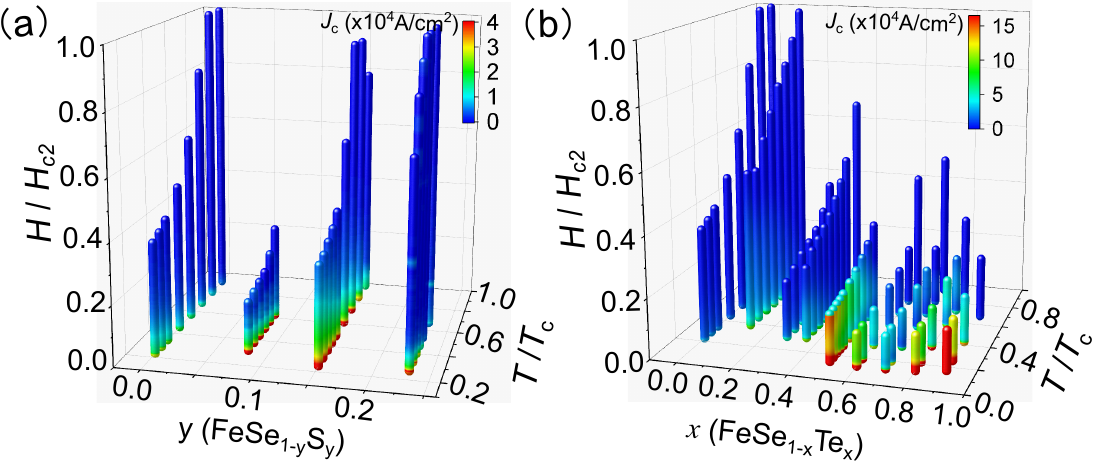}
	\caption{Critical current density $ J_{\rm{c}} $ for FeSe$ _{1-y} $S$ _{y} $ (a) and FeSe$ _{1-x} $Te$ _{x} $ (b) under normalized magnetic fields ($ H/H_{\rm {c2}} $) and temperatures ($ T/T_{\rm{c}} $). The magnitude of $ J_{\rm{c}} $ is indicated by the color scale. }\label{}
\end{figure*}

\end{document}